\documentclass[twocolumn,showpacs,nofootinbib]{revtex4}
\usepackage{epsf}
\usepackage{graphicx}
\usepackage{dcolumn}

\def\lsim{~\rlap{$<$}{\lower 1.0ex\hbox{$\sim$}}}
\def\gsim{~\rlap{$>$}{\lower 1.0ex\hbox{$\sim$}}}

\newcommand{\BE}{\begin{equation}}
\newcommand{\EE}{\end{equation}}
\newcommand{\BA}{\begin{eqnarray}}
\newcommand{\EA}{\end{eqnarray}}

\def\be{\begin{equation}}
\def\ee{\end{equation}}
\def\bea{\begin{eqnarray}}
\def\eea{\end{eqnarray}}

\def\la{\mathrel{\mathpalette\fun <}}
\def\ga{\mathrel{\mathpalette\fun >}}
\def\fun#1#2{\lower3.6pt\vbox{\baselineskip0pt\lineskip.9pt
        \ialign{$\mathsurround=0pt#1\hfill##\hfil$\crcr#2\crcr\sim\crcr}}}

\begin{document}
\input epsf
\renewcommand{\topfraction}{0.8}
\preprint{astro-ph/yymmnnn, \today}


\vspace{1cm}

\title{\bf\LARGE Probing Cosmology with Weak Lensing Peak Counts}

 \author{\bf Jan M. Kratochvil$^{1}$, Zolt\'an Haiman$^{2,1}$, Morgan May$^{3}$}

\affiliation{ {$^1$ Institute for Strings, Cosmology, and Astroparticle Physics (ISCAP), Columbia University, New York, NY 10027, USA}} 
\affiliation{ {$^2$ Department of Astronomy and Astrophysics, Columbia University, New York, NY 10027, USA}} 
\affiliation{ {$^3$ Physics Department, Brookhaven National Laboratory, Upton, NY 11973, USA}   }


{\begin{abstract} 

    We propose counting peaks in weak lensing (WL) maps, as a function
    of their height, to probe models of dark energy and to constrain
    cosmological parameters. Because peaks can be identified in
    two--dimensional WL maps directly, they can provide constraints
    that are free from potential selection effects and biases involved
    in identifying and determining the masses of galaxy clusters. We
    have run cosmological N--body simulations to produce WL
    convergence maps in three models with different constant values of
    the dark energy equation of state parameter, $w=-0.8, -1$, and
    $-1.2$, with a fixed normalization of the primordial power
    spectrum (corresponding to present--day normalizations of
    $\sigma_8=0.742,0.798$, and $0.839$, respectively).  By comparing
    the number of WL peaks in 8 convergence bins in the range of $-0.1
    < \kappa < 0.4$, in multiple realizations of a single simulated
    $3\times3$ degree field, we show that the first (last) pair of
    models can be distinguished at the 95\% (85\%) confidence level.
    A survey with depth and area (20,000 degree$^2$), comparable to
    those expected from LSST, should have a factor of $\approx 50$
    better parameter sensitivity.  We find that relatively
    low--amplitude peaks ($\kappa\sim 0.03$), which typically do not
    correspond to a single collapsed halo along the line of sight,
    account for most of this sensitivity.  We study a range of
    smoothing scales and source galaxy redshifts ($z_s$). With a fixed
    source galaxy density of 15 arcmin$^{-2}$, the best results are
    provided by the smallest scale we can reliably simulate, 1
    arcminute, and $z_s=2$ provides substantially better sensitivity
    than $z_s\leq 1.5$.

\end{abstract}}
\maketitle

\section{Introduction}\label{Introduction}

Weak gravitational lensing (WL) has emerged as one of the most
promising methods to constrain the parameters of both dark energy (DE)
and dark matter (DM) \cite{DETF}.  This method exploits the fact that
images of distant galaxies are distorted as their light is lensed by
inhomogeneities in the foreground gravitational potential.  DE and DM
alter the expansion history of the universe, as well as the rate at
which cosmic dark matter structures grow from small initial
perturbations, and thereby modify the statistical features of the WL
signal \cite{RefReview}.  Several deep astronomical surveys plan to
cover large regions ($\gsim10^3$ degree$^2$) of the sky in the future,
detecting millions of background galaxies, and measuring their shapes
precisely, in order to reveal the statistical features of the WL
distortion.  Examples include the Panoramic Survey Telescope And Rapid
Response System (Pan-STARRS) \cite{Kaiser:1999kka}, the Kilo-Degree
Survey (KIDS)\footnote{http://www.strw.leidenuniv.nl/~kuijken/KIDS},
the Dark Energy Survey
(DES)\footnote{https://www.darkenergysurvey.org}, and the survey by
the Large Synoptic Survey Telescope (LSST) \cite{Tyson:2002nh,
Tyson2002b} covering half of the sky.

Existing theoretical studies assessing the utility of such surveys to
probe the properties of DE and of DM can be roughly divided into two
types.  The first type analyzes the WL distortion primarily on large
scales, caused by fluctuations in the foreground mass distribution on
correspondingly large scales.  On these scales, the fluctuations, and
hence the WL distortion patterns, are linear or quasilinear, and can
be readily and reliably interpreted using existing semi--analytic
tools \citep{Kaiser92,JS97,WH99,WH02,DH02,AR04,A&D03,T&J04,S&K04}.
These studies (for example, those computing the two--dimensional power
spectrum of the shear) have, in fact, been pushed into the non--linear
regime, but the dependence of the non--linear power--spectrum on
cosmological parameters has not been verified in simulations, to an
accuracy necessary to take full advantage of the future surveys.

The second type analyzes the distortion on smaller scales, where it is
dominated by the nonlinear mass fluctuations.  In the limit of
extremely non--linear, high--$\sigma$ WL shear peaks, the individual
peaks tend to be caused by a single, discrete, massive collapsed
object -- a galaxy cluster -- in the
foreground~\citep{whiteprojection, Hamana:2003ts,Hennawi:2004ai}.  The
abundance evolution and spatial distribution of galaxy clusters in
different cosmologies can then be studied semi--analytically
(calibrated by simulations), and has been shown to place very strong
constraints on DE parameters, because it is exponentially sensitive to
the growth rate of dark matter fluctuations. Several studies have
explored constraints expected from samples of tens of thousands of
galaxy clusters in Sunyaev--Zeldovich and X--ray surveys
(e.g.~\citep{W&S98,H&M&H01,W&B&K02}), and from the still larger
samples of clusters expected to be detectable in large weak lensing
surveys \citep{SW04,Marian:2008fd,Chanta}.

The amount of statistical information in these two regimes appears to
be comparable \citep{Fang:2006dt}; furthermore, the information from
the two approaches is complementary, and using them in combination
significantly improves DE constraints compared to using either method
alone \citep{Fang:2006dt,Takada:2007}.

The chief difficulty with using galaxy clusters identified in WL maps,
however, is that the correspondence between peaks in WL maps and
collapsed clusters is not one--to--one. A significant fraction of WL
peaks are caused by chance alignments of multiple non--linear
structures along the line--of--sight. Conversely, a non--negligible
fraction of real clusters are missed because they do not show a
significant WL peak, due to a chance cancellation of their shear
signal by under-dense regions in the foreground.  In principle, the
statistical correspondence between WL peaks and clusters can be
precisely quantified ab--initio in numerical simulations, but in
practice, this remains a challenge to the required level of
accuracy. In particular, the completeness and purity of cluster
selection are of the order of $60-80\%$ for peaks selected to lie
above a $\sim 5\sigma$ convergence threshold, and both become
progressively worse for lower S/N peaks
\citep{whiteprojection,SW04,M&B06, Schirmer:2006ud, Dietrich:2007gi}.
Systematic selection errors are therefore likely to dominate over the
statistical constraints for the majority of a WL--selected cluster
sample.

A natural way to circumvent such selection effects is to study the
statistics of non--linear peaks in WL maps directly. This method,
however, does not lend itself to straightforward mathematical analysis
-- it requires numerical simulations and has been relatively much less
well--explored ~(e.g.~\citep{BJ&VW00,Wang:2008hi}).  In this paper, we
study WL peaks in simulated maps, and their usefulness to constrain DE
properties.  More specifically, we use a set of N-body simulations,
combined with ray-tracing, to create convergence maps for three
different values of the DE equation of state, $w=p/\rho$. We then
identify peaks in such maps, and tabulate them as a function of their
height.  We explicitly avoid a reference to clusters, cluster masses,
or mass functions -- we utilize only the directly observable height of
the maximum of each convergence peak. This method should therefore
provide one of the most robust cosmological probes, free from
potential selection effects and mass measurements biases affecting
cluster identifications.

For increasingly negative values of $w$, DE starts to dominate at
increasingly later cosmic epochs, and one expects an excess of high
peaks in such models, compared to those with less negative $w$.  In
this first paper, we perform a feasibility study, to asses whether
changes in $w$ indeed have a measurable effect on the number of
peaks. To this end, all cosmological parameters, other than $w$, are
held fixed at their fiducial values.\footnote{It is important to note
here, however, that we fix the primordial amplitude of the density
perturbations; this results in a $w$--dependent value of the
present--day normalization $\sigma_8$.  In the rest of this paper,
whenever we compare cosmologies with two different values of $w$, it
should be remembered, even when not explicitly stated, that these
two cosmologies also have different values of $\sigma_8$.} In a
follow--up paper, we will vary other cosmological parameters, thereby
exploring degeneracies between parameter combinations and obtaining
more realistic, marginalized constraints, on each parameter.

While the cosmological dependence of the cluster counts has been
studied extensively, it is not a priori clear whether the full catalog
of all WL peaks contains a comparable amount of information.  A large
number of peaks caused by large-scale structure, with no
$w$--dependence, would constitute pure noise, and could wash out the
theoretically known difference from peaks due to clusters.  A strong
$w$-dependence in the number of non--cluster peaks, in the sense
opposite to that of the cluster peaks, could be even more detrimental,
since this would explicitly cancel some of the signal contained in the
cluster counts.  Conversely, a strong $w$-dependence in the number of
non--cluster peaks, in the same sense as the cluster counts, could
significantly enhance the sensitivity. This remains a realistic
possibility, since the dependence of the number of peaks caused by
line of sight projections on cosmology has not yet been studied.

An intuitive expectation is that large-scale structure filaments are
expected to contribute more strongly to relatively low--height peaks,
and will be more prominent for source galaxies at higher redshift.
This is because most filaments are likely to line up only partially,
thus creating a weaker signal than those caused by massive collapsed
clusters, and because the probability to find large ($\sim 100$Mpc)
filaments that are lined up along the line of sight will increase
significantly at larger distances.  We perform our simulations at high
mass--resolution (down to less than $5\times10^{9}M_\odot$ per
particle), allowing us to resolve relatively low significance peaks in
our convergence maps, where such projections of large scale structure
are more common,. We also place the source galaxies out to redshifts
as high as $z_s=2$.

To keep the computational cost of the simulations reasonable, we
sacrifice the angular size of our simulated field of view, limiting
ourselves to creating WL maps covering $2.86\times2.86$ degrees (for
simplicity, we will hereafter refer to these as $3\times3$ degree
fields).

In these respects, our work differs from the recent related study by
\cite{Marian:2008fd}, which considered the projected mass function out
to a lower redshift (to $z=0.3-0.6$), with simulations at
significantly lower resolution, to demonstrate that the 2D projected
mass function obeys a scaling with $\Omega_m$ and $\sigma_8$ that is
similar to that of the three--dimensional halo mass function
(\cite{Marian:2008fd} also do not vary $w$). Our work shows that the
depth of the survey (at least to $z\sim2$) is vital for distinguishing
between models with different $w$, and that the main distinguishing
power comes from medium--height convergence peaks, with convergence
$\kappa\sim 0.03$ (on maps with 1 arcmin smoothing).

Another recent study, \cite{Wang:2008hi} explored probing cosmology
with a related, simple statistic: the high tail of the one--point
probability distribution of the convergence.  The total area above a
fixed convergence threshold includes a sum of the area of all peaks
lying above this threshold, and we therefore expect it to capture less
information than peak counts. Nevertheless, \cite{Wang:2008hi} have
found that this statistic can already place constraints on DE
properties, comparable to the traditional linear two point correlation
function of shear or to the limited use of the nonlinear information
from counting galaxy clusters \citep{Wang:2008hi}.  Our paper improves
on this preliminary study, by employing direct simulations, rather
than fitting functions, for the convergence and its dependence on
cosmology, and also by examining a potentially more powerful
statistic.

As our paper was being prepared for submission, we became aware of an
independent preprint, addressing the dependence of WL peaks on the
background cosmology \cite{Dietrich:2009jq}.  Their motivation is
essentially identical to our own, and the conclusions reached are
quite similar. However, the two studies are complementary, in that
\cite{Dietrich:2009jq} study the peak dependence in the $(\Omega_m,
\sigma_8)$ plane assuming a $\Lambda$CDM cosmology, while we vary the
dark energy equation of state parameter $w$ (with corresponding
changes in $\sigma_8$).  There are additional differences, such as the
specifications in the simulations that were used, and the statistical
analysis methods, which we briefly discuss further in \S~\ref{Other
Constraints} below.

The rest of this paper is organized as follows.
In \S~II, we describe our calculational procedure, including the set--up and analysis of the simulations.
In \S~III, we test our computations by comparing our numerical convergence power spectrum to semi-analytic predictions.
In \S~IV, we present and discuss our main results on the peak counts.
Finally, in \S~V, we summarize our main conclusions and the implications of this work.

\section{Methodology}
\label{Methodology}

This section describes the procedure to create the simulated weak
lensing maps, from generating the matter power spectrum with CAMB and
scaling it to the starting redshift of the N-body simulation, to
generating the initial conditions for the simulation, ray-tracing the
weak lensing maps from the simulation output, as well as
post--processing and analyzing these maps.

\subsection{N-body Simulations}\label{N-body}

For this study, we generated a series of cold dark matter (CDM) N-body
simulations with the code GADGET-2 \cite{Springel:2005mi}, which
include DM only; baryons and neutrinos are both assumed to be absent.
We have modified GADGET-2 for use with arbitrary equation of state
parameter $w$, but in this work, we restrict ourselves to constant
$w$. As the reference model, we chose a $\Lambda$CDM model with
parameters close to the best-fit values from WMAP5, i.e. with
cosmological constant $\Omega_\Lambda=0.74$, matter density parameter
$\Omega_m=0.26$, Hubble constant $h=0.01\times (H_0/{\rm km~s^{-1}
Mpc^{-1}})=0.72$. We fix the amplitude of curvature fluctuations
$\Delta_R^2=2.41\times10^{-9}$ at the pivot scale $k=0.002
\mathrm{Mpc}^{-1}$.  This results in different values of the
present--day normalization, $\sigma_8$, for the different dark energy
models.  In the three models we consider with $w=-0.8, -1$, and
$-1.2$, the values are $\sigma_8=0.742,0.798$, and $0.839$,
respectively.

For each of the three cosmologies, with all parameters other than $w$
and $\sigma_8$ fixed, we ran two independent N-body simulations, with
two different random realizations of the initial particle positions
and velocities. The simulations used $512^3$ particles, in a box with
a comoving size of $200h^{-1}\mathrm{Mpc}$, starting at redshift
$z=60$.  This corresponds to a mass resolution of
$m=4.3\times10^9M_\odot$ per dark matter particle. The
Plummer-equivalent gravitational softening of the simulations was set
to $\epsilon_{Pl}=7.5h^{-1}\mathrm{kpc}$ (comoving). Fixing the
softening length in physical (rather than comoving) units is not
warranted, since for a pure CDM simulation without gas, no further
physical scale enters.

The initial conditions (ICs) for the simulations were generated with
an IC generator kindly provided by Volker Springel. The matter power
spectrum at $z=60$, which is an input for the IC generator, was
created with the Einstein-Boltzmann code CAMB \cite{Lewis:1999bs}, a
modification/extension of CMBFAST \cite{Seljak:1996is}. GADGET-2 does
not follow the evolution of the radiation density, which, starting
with the correct power spectrum at $z=60$, would accumulate to a $\sim
2\%$ error in its normalization by $z=0$. To eliminate this small
bias, we used CAMB to produce the power spectrum at $z=0$ and scaled
it back to $z=60$, with our own growth factor code, with the radiation
component absent.  This leads to a small deviation from the true power
spectrum at the starting redshift, but the match will become better at
the lower redshifts which are more important for our analysis.

The output of the simulations consists of particle positions at
various redshifts between $z=0$ and $z=2$. Similar to choices in most
previous works (e.g. \cite{Hamana:2001vz}), the output redshifts were
chosen to span comoving intervals of $70h^{-1}\mathrm{Mpc}$ along the
line of sight in the $\Lambda$CDM cosmology.  For simplicity, the same
redshifts were used in the other cosmologies. This results in a small
difference in comoving separation between the planes in these
cosmologies, but this has negligible impact on our results.

Most simulations and all post--processing analysis were run on the
LSST Cluster at the Brookhaven National Laboratory, an 80-CPU Linux
cluster with 80~GB of memory. Some simulations were run on the Intel
64 Cluster Abe, with 9600~CPUs and 14.4~TB of memory at the National
Center for Supercomputing Applications (NCSA), a part of the NSF
TeraGrid facility.

\subsection{Weak Lensing Maps}\label{Maps}

We use our own codes to produce and analyze weak lensing maps from the
N-body simulation output. This subsection describes the procedures and
algorithms, as well as the testing of the new codes.

\subsubsection{Potential Planes}

As mentioned above, the simulation outputs are produced at every $70
h^{-1}$Mpc, which results in an $\approx 2/3$ overlap between
consecutive $200h^{-1}$Mpc cubes.  We prepare the N-body simulation
output for ray-tracing by truncating the snapshot cubes along the line
of sight in order to remove this overlap, and then project the
particle density of each cube on a 2D plane $p$ perpendicular to the
line of sight at the center of the plane, located at the same redshift
as the original cube. We use the triangular shaped cloud (TSC) scheme
\cite{Hockney-Eastwood} to convert the individual particle positions
in the simulations to a smoothed density field on the 2D plane.  The
projected density contrast field on plane $p$,
$\delta_p^{(2D)}(\vec{x})=\int_{y_{p-1}}^{y_p} dy
[\rho(\vec{x},y)/{\overline\rho}-1]$ is calculated on a 2D lattice of
$2048\times2048$ points (here $\vec{x}$ represents the coordinates in
the plane perpendicular to the line of sight; $y$ is the radial
coordinate, and $y_{p}$ is the position half--way between the $p^{\rm
th}$ and $(p+1)^{\rm th}$ planes). We then use the 2D Poisson equation
\BE
\nabla^2\Psi^p(\vec{x})=\frac{3\Omega_mH_0^2}{c^2}\delta_p^{(2D)}(\vec{x})
\EE
to calculate the 2D gravitational potential $\Psi^p$ on plane
$p$. This equation is solved easily in Fourier space, in particular
since the density contrast planes have periodic boundary
conditions. The resulting potential plane is then stored in FITS
format \cite{FITS} for further use by our ray-tracing code. This
multi-step approach has proven useful to save memory, and for
re--using intermediate steps of the lensing map construction.

\subsubsection{Ray-Tracing}

To create weak lensing maps from the N-body simulations, we implement
the two-dimensional ray-tracing algorithm described in
\cite{Hamana:2001vz}. We follow $2048\times2048$ light rays through a
stack of potential planes, which, as mentioned above, are spaced
$70h^{-1}\mathrm{Mpc}$ apart in the case of the $\Lambda$CDM
cosmology.  Starting at the observer, we follow the light rays
backward to the source galaxies. At each lens plane $n$, we calculate
the distortion tensor and the gravitational lensing deflection from
the equations:
\BA
\label{distortion tensor}
\mathbf{A}_n&=&\mathbf{I}-\sum_{p=1}^{n-1}\frac{f(\chi_p)f(\chi_n-\chi_p)}{a(\chi_p)f(\chi_n)}\mathbf{U}_p\mathbf{A}_p\\
\label{deflection angle}
\theta^n&=&\theta_1-\sum_{p=1}^{n-1}\frac{f(\chi_n-\chi_p)}{a(\chi_p)f(\chi_n)}\nabla_\bot\Psi^p,
\EA 
where $\mathbf{I}$ is the identity matrix, $f(\chi)$ and $a(\chi)$
denote the comoving angular diameter distance and the scale factor,
evaluated at the comoving coordinate distance $\chi$, and the
so--called optical tidal matrix,
\BE 
\mathbf{U}_p=\left(\begin{array}{cc} \Psi^p_{,11} &
\Psi^p_{,12} \\ \Psi^p_{,21} & \Psi^p_{,22} \end{array} \right) 
\EE
contains the second spatial derivatives of the gravitational potential
$\Psi$ on plane $p$. The derivatives are to be evaluated at the point
on plane $p$ intersected by the light ray. Starting with
$\mathbf{A}_0=\mathbf{I}$, equation~(\ref{distortion tensor}) can be
solved to find $\mathbf{A}_n$ for any $n$.

The physical weak lensing quantities, the convergence $\kappa$ and the
two components of the shear, $\gamma_1$ and $\gamma_2$, are extracted
from the distortion tensor via
\BE
\mathbf{A}=\left(\begin{array}{cc} 1-\kappa-\gamma_1 & -\gamma_2-\omega \\ -\gamma_2+\omega & 1-\kappa+\gamma_1 \end{array} \right).
\EE
The observable quantities, the magnification $\mu$ and the distortion
$\underline{\delta}$, are given by
\BE
\mu=\frac{1}{(1-\kappa)^2-\gamma^2}\hspace{2cm}\underline{\delta}=\frac{2\underline{g}}{1+|\underline{g}|^2},
\EE
where $\underline{g}=\underline{\gamma}/(1-\kappa)$ is the reduced shear. In the
weak lensing limit these relations simplify to $\mu=1+2\kappa$ and
$\underline{\delta}=2\underline{\gamma}$.

For simplicity, for the purpose of this paper, we will work directly
with the maps of the theoretically predicted convergence $\kappa$.  In
practice, in the absence of external size or magnification
information, neither the shear, nor the convergence is directly
observable, but only their combination, the reduced shear
$\underline{g}$. One can, in principle, convert one quantity to the
other; in practice, on small scales, where lensing surveys get much of
their constraining power, the errors introduced by this conversion
will not be negligible, and must be taken into account when extracting
cosmological parameters from actual data \cite{Dodelson:2005ir}.

\subsection{Source Galaxies, Smoothing and Noise}\label{Smoothing}

The point--particles in the N--body simulations represent lumps of
matter with a finite spatial distribution. While gravitational
softening is included in GADGET-2 to prevent unrealistically close
encounters, the simulations still follow the assembly of point
particles.  As a result, several additional types of smoothing are
necessary to produce weak lensing maps.  The TSC scheme, which is used
to place particles on the density contrast planes, generates smooth
two--dimensional density contrast distributions, thus avoiding
unnaturally large deflections when a light ray would otherwise pass
close to an individual particle. In the ray-tracing step, as light
rays pass through arbitrary points on the potential planes, we employ
the same TSC kernel to compute the potential at the ray intersection
points.

Since the orientation of the disks of the galaxies with respect to the
observer is unknown, this intrinsic ellipticity appears as noise in
the maps. Following the interpretation of \cite{VanWaerbeke:1999wv} in
\cite{Fang:2006dt}, we take the redshift-dependent r.m.s. of the noise
in one component of the shear to be \cite{S&K04}
\BE\label{gamma noise}
\sigma_\gamma(z)=0.15+0.035z
\EE
and perform the calculation analogously to \cite{VanWaerbeke:1999wv}
but with a square top-hat function representing our pixels. The
resulting convergence noise correlation function for our pixelized map
where $\vec{x}$ and $\vec{y}$ are the discrete 2D pixel coordinates
is:
\BE\label{noise correlation} 
\langle \kappa_{\rm noise}(\vec{x})\kappa_{\rm noise}(\vec{y})\rangle=\frac{\sigma_\gamma^2}{n_{gal}A}\delta_{\vec{x}\vec{y}},
\EE
where $\delta_{\vec{x}\vec{y}}$ is the Kronecker symbol, $n_{gal}$ is
the surface density of source galaxies, and $A$ is the solid angle of
a pixel. For simplicity, in our analysis we assume that the source
galaxies are located on one of three source planes, at fixed redshifts
($z_s=1, 1.5, 2$), with $n_{gal}= 15~{\rm arcmin}^{-2}$ on each source
plane. In combination, this would yield a total surface density of
$n_{gal}= 45~{\rm arcmin}^{-2}$. For comparison, we note that this is
larger than the value $\approx 30~{\rm arcmin}^{-2}$ (with a
redshift--distribution that peaks at $z_s\sim 1$), expected in LSST.
However, we do not have sufficient simulation data to perform accurate
tomography, and therefore, our results below utilize only a single
source plane, with a conservative number of $n_{gal}= 15~{\rm
arcmin}^{-2}$, rather than the full $n_{gal}= 45~{\rm arcmin}^{-2}$
(see below).  Thus, we add noise to the convergence in each pixel,
drawn from a Gaussian distribution with the variance given in
equation~(\ref{noise correlation}).  Note that the noise between
different pixels is uncorrelated.

In order to investigate the information content in peaks defined on
different scales, once noise in each pixel is added, we apply an
additional level of smoothing, with a finite 2D Gaussian kernel, on
scales between $\theta_G = 0.5-10$ arcminutes (see
Sec.~\ref{Comparison} for exact values).  The Gaussian kernel has the
nice property of being decomposable, thus in practice, we can
equivalently make two passes through the map with 1D Gaussian kernels,
oriented once in each of the two orthogonal dimensions of the
map. This saves computation time, especially for large smoothing
scales.

The noise in the convergence $\sigma^2_{noise}$ introduced by
intrinsic ellipticity, after the Gaussian smoothing, becomes
\BE\label{noise}
\sigma_{noise}^2=\frac{\langle\sigma_\gamma^2\rangle}{2\pi\theta_G^2n_{gal}}
\EE
which is the equivalent of (\ref{noise correlation}), but with a
Gaussian smoothing instead of a top-hat function
\cite{VanWaerbeke:1999wv}, and depends on the redshift distribution of
the source galaxies, as well as on the smoothing scale $\theta_G$
applied to the maps. In general, the noise level should be computed as
the weighted average of the redshift--dependent shear noise
(\ref{gamma noise}) over the redshift distribution of the source
galaxies:
\BE
\langle\sigma_\gamma^2\rangle=\frac{1}{n_{gal}}\int_0^\infty\sigma_\gamma^2(z)\frac{dn}{dz}dz.
\EE
Since in our case, for simplicity, we assume that the galaxies are
located on three discrete source planes, $dn/dz$ is simply a delta
function, at $z_s=1, 1.5$, or at $z_s=2$ (or a sum of delta functions,
for the purpose of redshift tomography).

\vspace{\baselineskip}
\subsection{Statistical Comparisons of Convergence Maps}\label{Comparison}

\begin{widetext}

\begin{figure}[htp]
\centering
\includegraphics[width=8 cm]{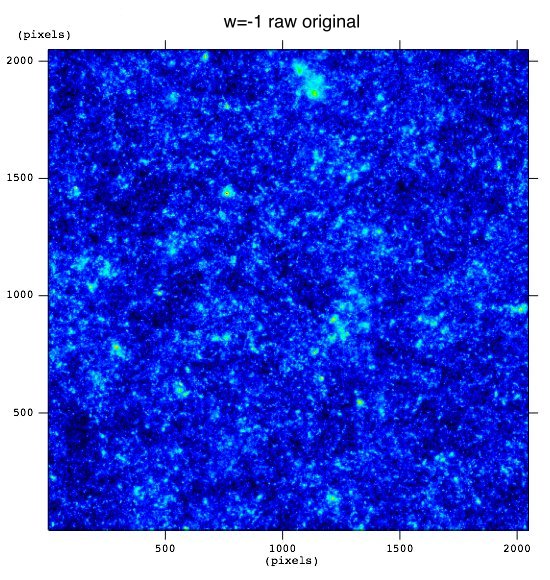} \includegraphics[width=8 cm]{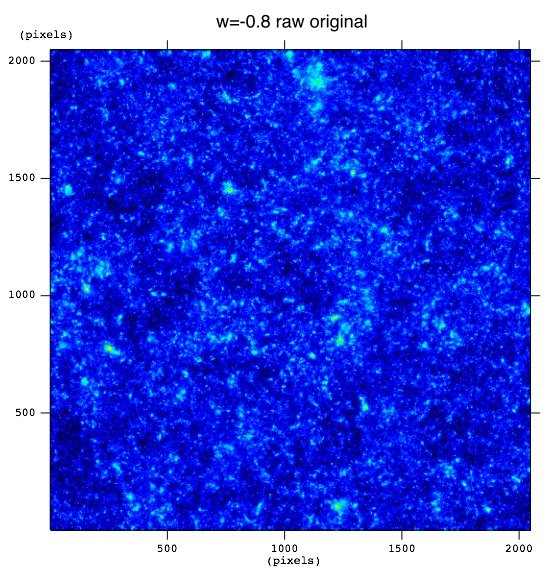}\\
\includegraphics[width=8 cm]{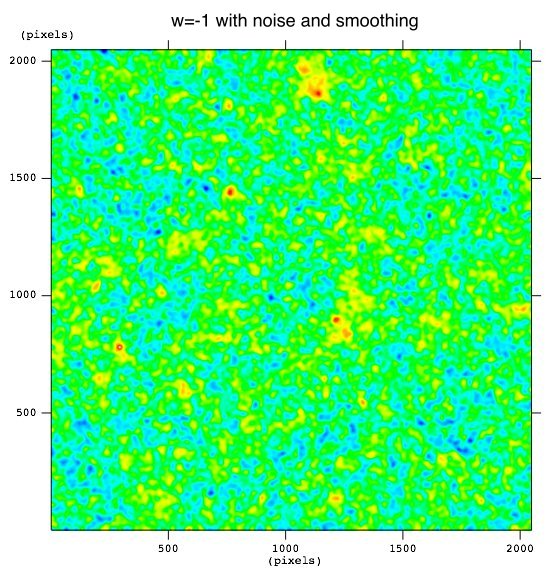} \includegraphics[width=8 cm]{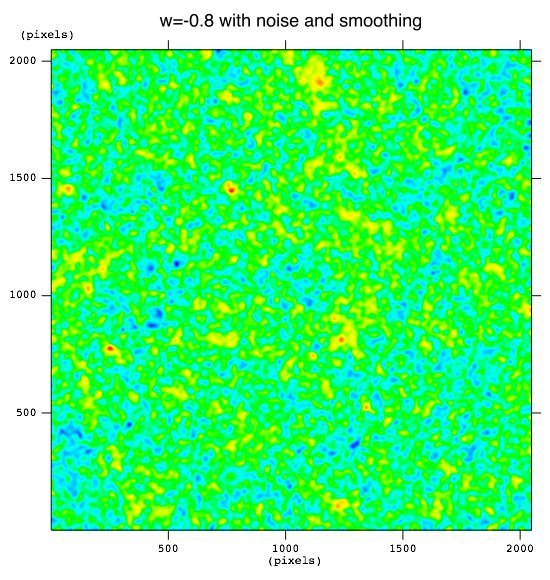}
\hfill
\caption[]{\textit{Examples of $3\times3$ degree convergence maps from
our $512^3$ particle N-body simulations in two different
cosmologies. The left column shows maps for $w=-1$, the right column
for $w=-0.8$. The top row shows the raw maps, with no smoothing or
noise; the bottom row shows the maps after the addition of random
ellipticity noise and 1 arcminute smoothing. The panels with $w=-1$
exhibit visibly more structure at high convergence (yellow and red in
the bottom panels); voids (regions of low convergence, shown in blue)
also appear more numerous in this cosmology.}}\label{fig:maps}
\end{figure}

\end{widetext}

The main goal of this paper is to discern between different
cosmologies by counting the convergence peaks. Since the maps have
been smoothed, as mentioned above, it is sufficient to identify peaks
in the maps as pixels which correspond to local maxima, in the sense
that all eight of their neighboring pixels have a lower value.  For
illustration, we show an example of the convergence maps generated in
two of our cosmologies, before and after the addition of noise and
smoothing, in Figure~\ref{fig:maps}.  The panels with $w=-1$ exhibit
visibly more structure at high convergence, consistent with the
expectation that DE starts to dominate later in this model, allowing
more time for growth.

Once the peaks are identified, their heights are recorded in a
histogram. This histogram is divided into a reasonable number of bins
(the choice in the number of bins will be discussed below), and a
vector of values is formed from these bins.  We will also attempt to
utilize the source galaxy redshifts for tomography, and to combine
different smoothing scales, so that each peak is further associated
with a value of $z_s$ and $\theta_G$.  Finally, peak identification is
repeated in each realization of a given cosmology.  We therefore
denote the number of peaks in bin $i$ with
$N_{i,z_s,\theta_G}^{(f;r)}$, where $f$ and $r$ indicate cosmology and
realization, respectively, $z_s$ indicates the redshift of the sources
and $\theta_G$ is the smoothing scale applied to the final map. To
simplify notation we hereafter drop the last two lower indices and
merge them into the single index $i$.

In order to assess the statistical significance of any difference in
peak counts, we construct the covariance matrix $C^{(f)}$ for
cosmology $f$ of the counts in the different height-, redshift--, and
smoothing--scale bins $i$ and $j$,
\BE\label{Covariance Matrix}
C_{ij}^{(f)}=\frac{1}{R-1}\sum_{r=1}^R (N_i^{(f;r)}-\overline N_i^{(f)})(N_j^{(f;r)}-\overline N_j^{(f)}),
\EE
where $\overline N_i^{(f)}=R^{-1}\sum_{r=1}^R {N}_i^{(f;r)}$ is the
average number of peaks in bin $i$ in the fiducial model $f$, and the
summation in equation~(\ref{Covariance Matrix}) is over the total
number of realizations $R$ constructed for every cosmology.

We use 200 such realizations. However, as noted above, we have run
only two strictly independent realizations of each cosmology, i.e.,
two different simulations with different realizations of the initial
conditions.  Thus we produce additional pseudo-independent
realizations by the standard procedure of randomly rotating, shifting,
and slicing the data in the simulation output cubes at each redshift,
prior to generating the 2D density planes.  Our box--size,
$200h^{-1}$Mpc is much larger than the few comoving Mpc correlation
length of large--scale structures.  As mentioned above, the data cubes
are truncated at each output, so that we utilize only a radial slice
covering about 1/3rd of its length; the $2.86\times2.86$ degree field
furthermore utilizes only a fraction of this slice.  As a result,
at intermediate redshifts,
where the lensing kernel peaks, each realization utilizes about a
twelfth of the volume of the snapshot cube, allowing us to access, in each N--body run, 
 effectively $\sim 12$ truly independent realizations of
each lens plane. At the very highest redshift, the
situation becomes worse, with up to a third of the volume of the data
cubes used. However, due to the distance--dependence of the lensing
kernel, the planes near $z_s=2$, with these heavily sampled boxes, do
not contribute much to the final lensing signal.  The random rotations
and shifts will create many more than 12 essentially independent
random realizations of the incidental line--ups of multiple small
clusters and filaments.  On the other hand, the very largest clusters
-- which may create a substantial lensing peak by themselves -- will
tend to reappear in multiple realizations, and could lead to an
underestimation of the variance.  Fortunately, our main results come
from medium--height peaks, which typically are not caused by a single
cluster, so that any such underestimation at the high--convergence end
contributes relatively little to our final distinguishing power.

We use the (inverse of) the covariance matrix to compute a value of
$\chi^2_{f^\prime}(r)$ for each realization $r$ of a ``test''
cosmology, labeled by $f^\prime$, against the expectations in the
fiducial cosmology, labeled by $f$, 
\BA\label{chi2}
\chi^2_{f^\prime}(r)&=&\mathbf{dN}^{(f^\prime;r)}(C^{(f)})^{-1}\mathbf{dN}^{(f^\prime;r)}\\&=&\sum_{ij}dN_i^{(f^\prime;r)}(C^{(f)})_{ij}^{-1}dN_j^{(f^\prime;r)},
\EA 
where $dN_i^{(f^\prime;r)}\equiv N_i^{(f^\prime;r)}-\overline N_i^{(f)}$ is
the difference between the number of peaks in bin $i$ in realization
$r$ of the test cosmology, and the average number of peaks in the same
bin in the fiducial cosmology.

\begin{widetext}

\begin{figure}[htp]
\centering
\includegraphics[width=8 cm]{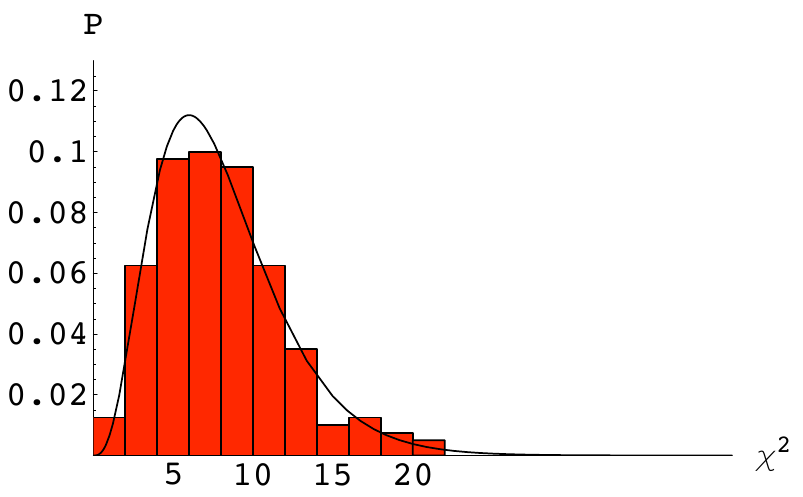} \includegraphics[width=8 cm]{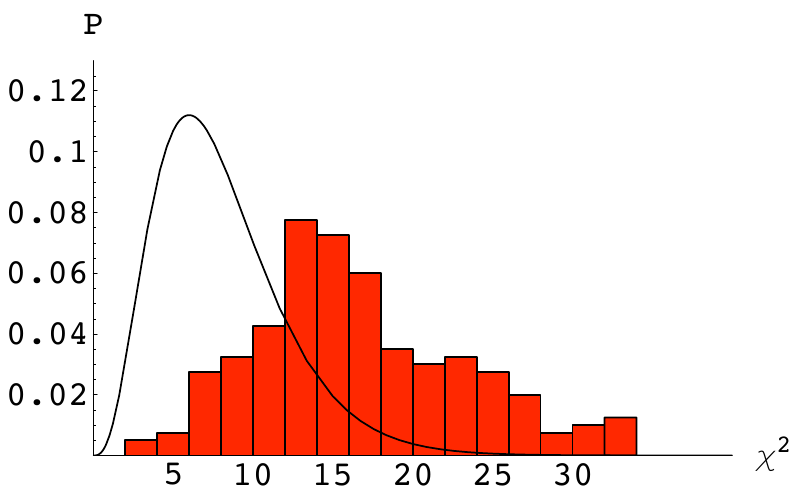}\\
\includegraphics[width=8 cm]{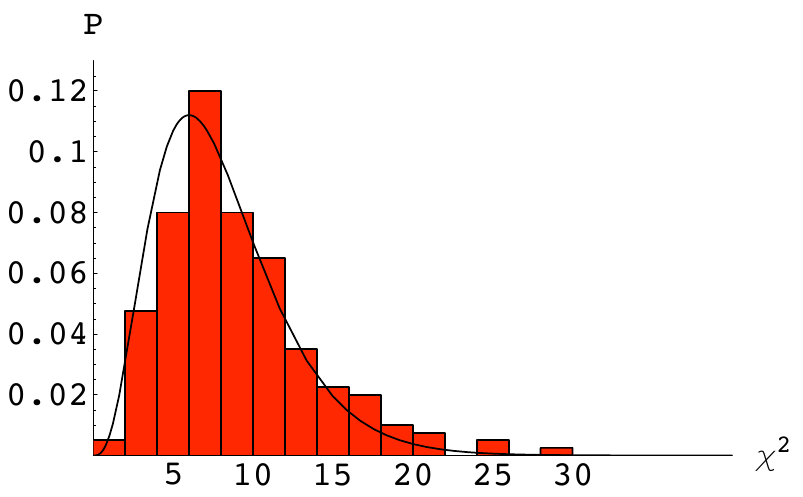} \includegraphics[width=8 cm]{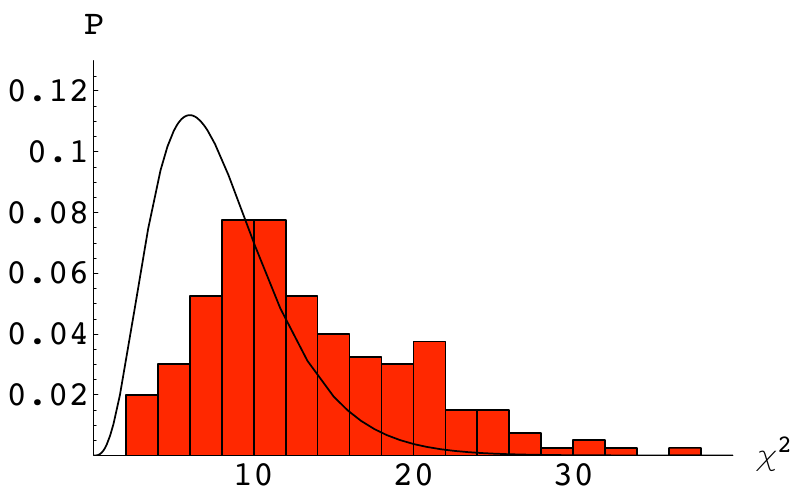}
\hfill
\caption[]{\textit{Normalized probability distribution of $\chi^2$,
computed by comparing the peak counts in individual $3\times3$ degree
convergence fields to the ensemble average in the fiducial $w=-1$
model. Peaks were divided into 8 bins by height. The source galaxies
were placed at redshift $z_s=2$, and a Gaussian smoothing of
$\theta_G=1$ arcmin was applied after the addition of ellipticity
noise. In the top left panel, the realizations were drawn from the
$w=-1$ simulation, the average of which was used to construct the
fiducial model. The black curve shows a genuine $\chi^2$ distribution
with $n=8$ degrees of freedom. In the bottom left panel, the
realizations were again drawn from a $w=-1$ model, but from a
simulation with different random initial conditions, illustrating that
these have a negligible effect on the result for most of the
distribution, and testing the accuracy of our covariance matrix.  The
two right panels show $\chi^2$ distributions in different cosmologies,
with $w=-0.8$ (top right panel), and $w=-1.2$ (bottom right
panel). The $w=-0.8$ and $\Lambda$CDM cosmologies are distinguished at
the 68\% (95\%) confidence level in 84\% (39\%) of the realizations;
the $w=-1.2$ cosmology is distinguishable from $\Lambda$CDM at the
same confidence in 61\% (24\%) of the realizations.}}\label{fig:chi2}
\end{figure}

\end{widetext}

The resulting probability distributions for $\chi^2_{f^\prime}(r)$,
using 200 realizations of each cosmology, fixed source redshift
$z_s=2$ and smoothing scale $\theta_G=1'$, and 8 peak height bins, is
shown in Figure~\ref{fig:chi2}.  In the top left panel, the
realizations were drawn from the simulation of the fiducial $w=-1$
model itself, which was used to compute the mean number of peaks
$\overline N_i^{(f)}$. Our simulated $\chi^2$ distribution is very
close to a genuine $\chi^2$ distribution with $n=8$ degrees of freedom
(shown by the black curve). In the bottom left panel, the realizations
were again drawn from the $w=-1$ cosmological model, but from the
simulation with an independent realization of the initial conditions.
The good agreement with the top left panel tests the accuracy of our
covariance matrix (in particular, if the covariance had been
underestimated, then this second realization would have produced
larger $\chi^2$ values; see Appendix~\ref{Initial Conditions}).  The
two right panels show $\chi^2$ distributions computed in the
realizations of different cosmologies, with $w=-0.8$ (top right
panel), and $w=-1.2$ (bottom right panel). The $w=-0.8$ and
$\Lambda$CDM cosmologies are distinguished at the 68\% (95\%)
confidence level in 84\% (39\%) of the realizations; the $w=-1.2$
cosmology is distinguishable from $\Lambda$CDM at the same confidence
in 61\% (24\%) of the realizations.

For reference, the above confidence levels can be compared to a
simpler estimate of the statistical distinction between the two pairs
cosmologies, based on the value of $\Delta\chi^2_{f^\prime,f}$,
computed using the mean number of peaks in each bin (averaged over all
realizations),
\BA\label{meanchi2}
\chi^2_{f^\prime,f}&=&\mathbf{dN}^{(f^\prime,f)}(C^{(f)})^{-1}\mathbf{dN}^{(f^\prime,f)}\\&=&\sum_{ij}dN_i^{(f^\prime,f)}(C^{(f)})_{ij}^{-1}dN_j^{(f^\prime,f)},
\EA 
where $dN_i^{(f^\prime,f)}\equiv \overline N_i^{(f^\prime)}-\overline
N_i^{(f)}$ is the difference between the average number of peaks in
bin $i$ in cosmology $f^\prime$ and in cosmology $f$.  We find
$\Delta\chi^2=8.35$ between $w=-0.8$ and $w=-1$ (when then same
realization of the initial conditions are used in both cosmologies)
and $\Delta\chi^2=9.54$ (when different realizations are used).  The
covariance matrix is calculated from the $w=-1$ cosmology in both
cases.  Assuming that the likelihood distributions are Gaussians,
these values would correspond to a $2.8\sigma$ and $3\sigma$
difference between the two cosmologies.  This is in good agreement
with the 95\% exclusion of the mean of the $\chi^2$ distribution we
find from the individual realizations in the $w=-0.8$ model --
indicating that interpreting the $\Delta\chi^2$ directly as a
probability would only slightly overestimate the statistical
difference between the two cosmologies. The corresponding value
between the $w=-1.2$ and $w=-1$ models is $\Delta\chi2=4.65$ (with the
same realization of the initial conditions), again confirming that the
true likelihood is only somewhat weaker than this simple estimate for
the exclusion of the $w=-1.2$ model.

In the above analysis, we have chosen convergence bins whose width was
varied, so that they contain approximately equal number of counts
($\approx 350$ peaks/bin at 1 arcmin smoothing).  More specifically,
the bin boundaries are located at
$\kappa= -0.084, 0.0022, 0.015, 0.024, 0.032, 0.041, 0.052, 0.068,
0.41$.  While equal--width bins would perhaps be a more natural
choice, the number of peaks falls off exponentially with peak height,
and this would have the undesirable result of having many almost empty
bins, leading to large $\chi^2$ values excessively often (overestimate
of improbability of unlikely events), and strong deviations from a
true $\chi^2$ distribution (since the peak counts do not have a
Gaussian distribution, especially for bins with low average counts).

\section{Weak Lensing Power Spectrum}
\label{Testing}

\begin{figure}[htp]
\centering
\includegraphics[width=8 cm]{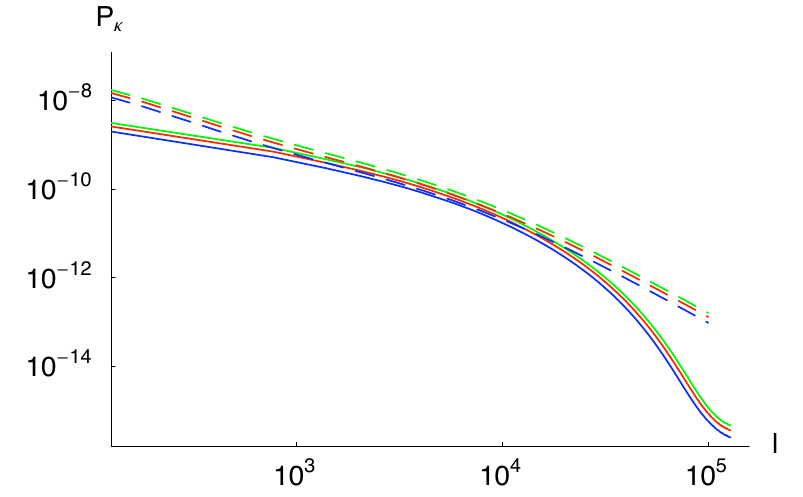} 
\hfill
\caption[]{\textit{Convergence power spectrum $P(\ell)$, as a function
of $\ell=2\pi/\theta$, derived in the three cosmologies (solid blue:
$w=-0.8$, solid red: $w=-1$, solid green: $w=-1.2$) from the raw
un-smoothed $3\times3$ degree maps, without ellipticity noise.  The
simulated power spectra have been averaged over 200 realizations in
each cosmology.  Also shown are the corresponding semi--analytical
predictions based on the matter power spectrum of \cite{Smith} and
using the Limber approximation \cite{Limber} (dashed curves).  The
drop around $\ell\sim20,000$, which corresponds to $\sim 1$ arcmin, is
due to the finite resolution of the simulations. The power is also
suppressed on large scales, because of finite box size. Note that the
ratios of power spectra in the three cosmologies are accurate in the
simulations over a wider range than the absolute
power.}}\label{fig:Spectra}
\end{figure}

Because our pipeline to produce the WL maps is newly assembled,
GADGET-2 was modified to allow $w\neq -1$, and the ray-tracing and WL
map analysis codes were newly written, we performed a consistency
check, by comparing the power spectrum of the convergence in our maps
with the power spectrum derived from semi--analytical
predictions. This comparison also determines the range of scales which
is adequately represented by our simulations due to resolution and box
size. In Figure~\ref{fig:Spectra}, we show the power spectrum of
convergence, averaged over 200 realizations of our simulations (solid
curves), together with the theoretical expectations (dashed curves).
The latter were obtained by direct line--of--sight integration using
the Limber approximation \cite{Limber}, and using fitting formulas for
the nonlinear 3D matter power spectrum from \cite{Smith}.  From top to
bottom, the three sets of curves correspond to $w=-1.2$ (green
curves), $w=-1$ (red curves), and $w=-0.8$ (blue curves).

We find significant deviations from the theoretical power spectra on
both small and large scales, as expected. On small scales, with $\ell
\ga 20,000$, we are missing power because of the finite resolution on
the mass planes and on the ray-tracing grid. On large scales, with
$\ell \la 400$, the spectrum is suppressed because of the finite size
of the simulation box.  These two wave numbers thus set the range,
from $\sim1$ arcmin to $\sim 1$ degree, on which the absolute value of
the converge power spectrum is accurately captured.

While we can not trust our results beyond these limits in the
subsequent analysis, it should be noted that the {\em ratio} of the
power spectra in different cosmologies is preserved accurately, even
where the simulated power spectra start to deviate from the
theoretical predictions. Since we are mainly interested in the
comparison between the cosmologies, this makes our results even more
robust.

Another consistency check we perform is a comparison between the
number of peaks above a certain threshold to the number of clusters
creating lensing peaks above that same threshold. This comparison will
presented in Sec.~\ref{Peak Numbers} below.

\section{Results and Discussion}
\label{Results}

We have found several important quantitative and qualitative results
in our study.  We first discuss the difference in peak counts between
the cosmologies -- in the total number of peaks, as well as in the
number of peaks in individual convergence bins. We then present our
inferred sensitivity to distinguish these cosmologies, and extrapolate
this sensitivity from our $3\times3$ degree field to an LSST--like
survey. We discuss the validity of such an extrapolation by studying
the variance in peak counts as a function of angular scale. In the
next subsection, we comment on correlations between peaks in different
bins, with source galaxies at different redshifts, and with different
angular smoothing scales, as well as cross-correlations among
these. Finally, we briefly contrast our results to other predictions
for an LSST--like WL survey, namely constraints from the weak lensing
shear power spectrum, the PDF of $\kappa$ found in \cite{Wang:2008hi},
as well as cluster number counts.

\subsection{Peak Counts}\label{Peak Numbers}

\begin{figure}[htp]
\centering
\includegraphics[width=8 cm]{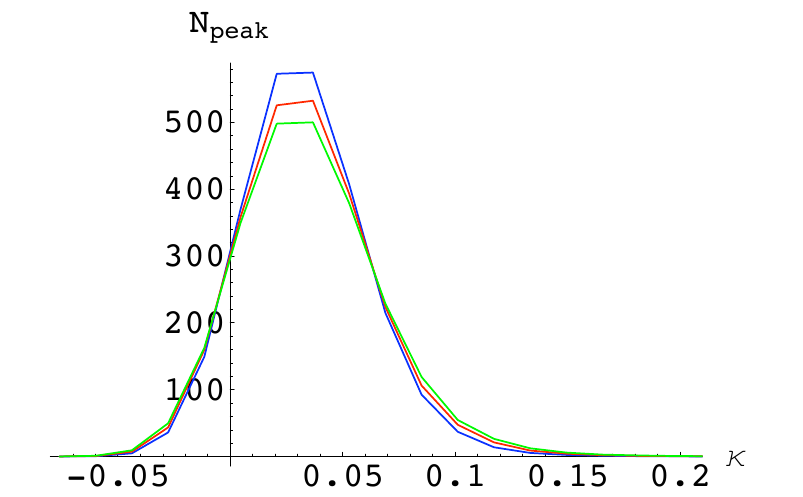}\\ \ \\
\hfill
\caption[]{\textit{Number of convergence peaks in intervals [$\kappa$,
$\kappa+\Delta\kappa$], with $\Delta\kappa=0.0161$ in $3\times3$
degree fields with source galaxy redshift $z_s=2$, 1 arcmin smoothing,
and intrinsic ellipticity noise added. From top to bottom (at the peak
of the distribution), the three curves correspond to $w=-0.8$ (blue),
$w=-1$ (red), and $w=-1.2$ (green). The total number of peaks is 2477,
2432, and 2401, respectively. Notice that the lower distributions have
wider wings.  }}\label{fig:N_Peak}

\end{figure}

The fundamental quantity we are interested in is the number of peaks
in different convergence bins. This quantity is shown in
Figure~\ref{fig:N_Peak} in the three different cosmologies, derived
from convergence maps with a single source galaxy plane at $z_s=2$,
after the addition of ellipticity noise and Gaussian smoothing with
$\theta_G=1$ arcmin.  From top to bottom (at the peak of the
distribution), the three curves correspond to $w=-0.8$ (blue curves),
$w=-1$ (red curves), and $w=-1.2$ (green curves).

As this figure demonstrates, the distribution of peak heights becomes
narrower with increasing $w$.

The total number of convergence peaks in the noisy, smoothed maps is
$N_{total}=2432\pm30$, $N_{total}=2477\pm30$, and $N_{total}=2401\pm
34$ for $w=-1$, $w=-0.8$, and $w=-1.2$, respectively. Formally, this
is a $1.5\sigma$ difference for $w=-0.8$ and a somewhat weaker
difference of $1\sigma$ for $w=-1.2$. The same numbers for the
smoothed noiseless maps (still with 1 arcmin smoothing) are
$N_{total}=1646\pm 24$, $N_{total}=1654\pm 28$, and $N_{total}=1644\pm
24$, respectively.  Interestingly, the addition of ellipticity noise
boosts both the total number of peaks (by $\sim 50\%$; note that a
similar increase was observed by \cite{Hamana:2003ts}), and the
significance in the difference in peak counts between pairs of
cosmologies.  Apparently, the number of new peaks introduced by the
presence of noise is, itself, $w$--dependent, in a way that helps in
distinguishing the two cosmologies.

While this result may first sound surprising, the addition of noise is
known boost the discriminating power of related observables. For
example, scatter in the mass-observable relation generally increases
the number of detectable clusters and tightens constraints from
cluster number counts (e.g \cite{H&M&H01}). In our case, one can image
that the number of new peaks on 1 arcmin scales, at a given peak
height, introduced by the presence of noise, depends on the amplitude
of the $\kappa$--fluctuations on larger angular scales, which, itself,
is cosmology--dependent.  The precise origin of this result will be
investigated in a follow-up paper.

Most of the $\sim 2,500$ peaks, of course, have a low amplitude. In
order to compare our results with previous work, it is of interest to
consider the number of peaks at higher significance.  As an example,
above the commonly used threshold of $\kappa_G=4.5\sigma_{noise}$ (for
a source galaxy density of $n_{gal}=15$ arcmin$^{-2}$ at redshift
$z_s=2$, the value of this threshold corresponds to $\kappa_G=0.102$
[eq.~\ref{noise}]), the average number of peaks we identify in our 200
pseudo-independent realizations is
$N_{\mathrm{peaks}>\kappa_G}=47.4\pm11.8$,
$N_{\mathrm{peaks}>\kappa_G}=36.6\pm10.4$, and
$N_{\mathrm{peaks}>\kappa_G}=54.5\pm 11.4$ for $w=-1$, $w=-0.8$ and
$w=-1.2$ respectively, which represent $\approx 1\sigma$ differences
between the $w=-1$ and the $w\neq 1$ models.  These numbers are with
ellipticity noise included. Without ellipticity noise, the number of
peaks above the same threshold is
$N_{\mathrm{peaks}>\kappa_G}=25.9\pm7.3$,
$N_{\mathrm{peaks}>\kappa_G}=14.8\pm5.2$, and
$N_{\mathrm{peaks}>\kappa_G}=37.0\pm9.5$, which, again, represent a
$1\sigma$ differences.  Although the number of these rare
high--$\sigma$ peaks is also boosted by the presence of noise,
apparently this boost does not enhance the distinguishing power
between cosmologies.

It is also useful to compare the number of peaks to the estimated
number of clusters causing a convergence signal above the same
threshold $\kappa_G$. To be closer to predictions in earlier work, for
this comparison, we take $n_{gal}=30$ arcmin$^{-2}$, resulting in a
threshold of $\kappa_G=0.072$ for 1~arcmin smoothing. In our noiseless
maps, we obtain $N_{\mathrm{peaks}>\kappa_G}=99.7\pm20.0$ and
$N_{\mathrm{peaks}>\kappa_G}=64.6\pm14.3$ for $w=-1$ and $w=-0.8$
respectively. To find the corresponding number of clusters, we repeat
the calculation in \cite{Fang:2006dt}, which is based on the fitting
formula for the DM halo mass function from \cite{Jenkins}, and assumes
clusters follow the Navarro-Frenk-White profile \cite{NFW}, but we
adjust the cosmological parameters (including the $w$--dependent value
of $\sigma_8$) to the values adopted here. We find the number of
clusters is $N_{\mathrm{clusters}>\kappa_G}=48.8\pm13$, and
$N_{\mathrm{clusters}>\kappa_G}=31.7\pm11$ for $w=-1$ and $w=-0.8$,
respectively (the errors bars here are approximately twice the
$\sqrt{N}$ Poisson errors, enhanced by cosmic variance, following
\cite{HuKravtsov}).  Apparently, there are approximately twice as many
convergence peaks in the maps without ellipticity noise as there are
clusters, indicating a substantial number of projections. This is
roughly consistent with the results of \citep{whiteprojection,
Hamana:2003ts,Hennawi:2004ai}, which self--consistently identify halos
in their simulations to quantify this correspondence more reliably,
although those works find a somewhat closer correspondence between
clusters and $4.5\sigma$ peaks. Adding ellipticity noise increases the
number of peaks by another factor of $\sim$two, resulting in plenty of
convergence peaks that are not due to a single cluster, which can
carry additional information about the cosmology.

An important point to note here is that, as we emphasized earlier, we
keep the primordial amplitude fixed as we vary $w$, which leads to a
different $\sigma_8$ for each $w$.  The difference in the average
number of clusters (48.8 versus 31.7) quoted above arises essentially
entirely from this difference in $\sigma_8$. We expect that the
difference in peaks counts will similarly be driven primarily by
$\sigma_8$.  Tomography should break some of the degeneracy between
$\sigma_8$ and $w$; this is an issue that will be clarified and
explored in our follow-up work.

\subsection{Three Peak Types}\label{peak types}

\begin{figure}[htp]
\centering
\includegraphics[width=8 cm]{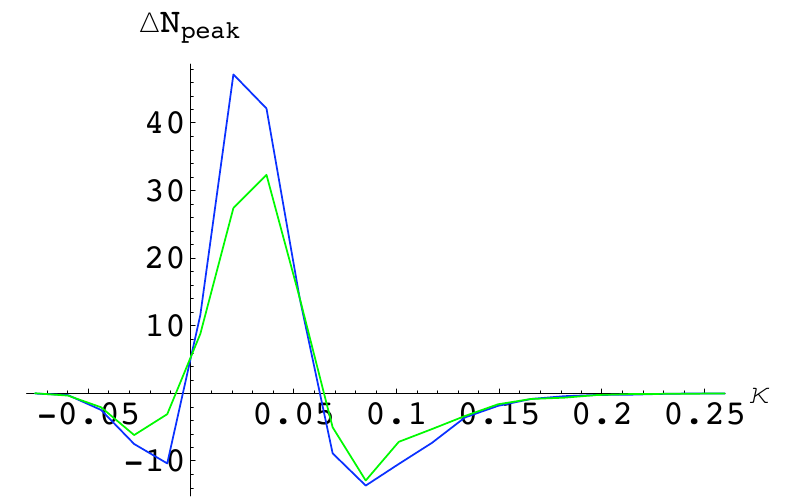}\\ \ \\
\includegraphics[width=8 cm]{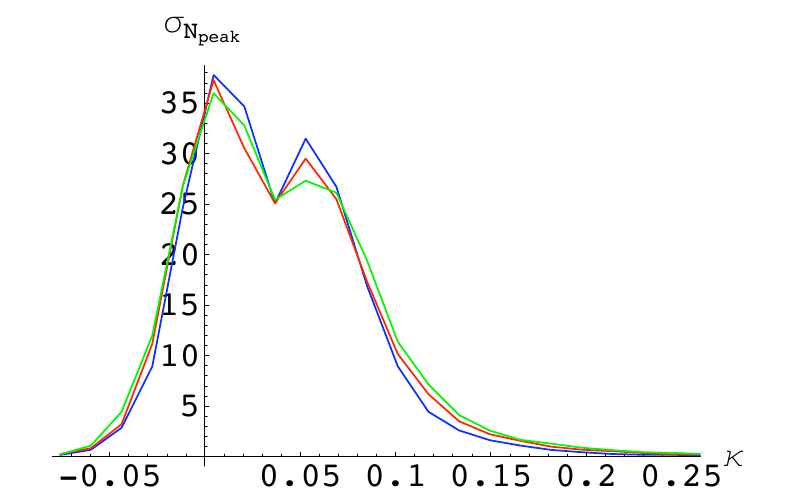}\\ \ \\
\includegraphics[width=8 cm]{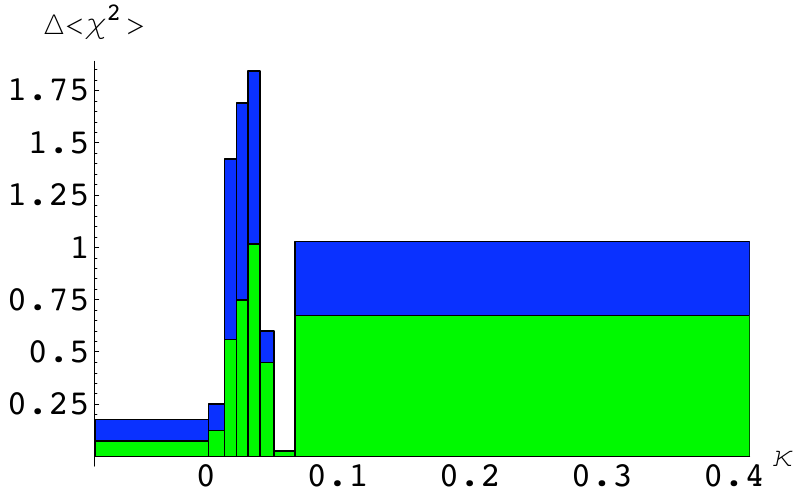}
\hfill
\caption[]{\textit{{\bf Top Panel}: the difference in average peak
numbers in each convergence bin is plotted as a function of
convergence $\kappa$. The $\Lambda$CDM cosmology is subtracted from
$w=-0.8$ (blue) and $w=-1.2$ is subtracted from $\Lambda$CDM (green)
in this plot. The height distribution of convergence peaks in
$\Lambda$CDM is broader, while in $w=-0.8$ it is narrower, and more
sharply peaked (and analogously for $w=-1.2$ vs.\ $\Lambda$CDM). The
numbers plotted are averaged over 200 realizations of a $3\times3$
degree field, with source galaxies at $z_s=2$, 1-arcminute smoothing
and the peak heights divided into 50 equal--width convergence bins.
{\bf Middle Panel}: Same as the upper panel, but showing the standard
deviation in the number of peaks each bin.
{\bf Bottom Panel}: Contribution to the total $\langle\chi\rangle^2$
from the peak number differences in each of the eight convergence bins
used in our $\chi^2$ calculation. Note that the bins have unequal
width, chosen such that each bin contains approximately the same
number of peaks (see text). Note that this panel neglects correlations
between bins, in order to illustrate the $\kappa$-dependence. There is
a small contribution from low peaks, a strong one from medium peaks,
and a sizable one from high peaks.  }}\label{fig:Delta_Peak}
\end{figure}

It is instructive to next take a more detailed look at how the number
of convergence peaks depends on the peak height, and, in particular,
how the counts in different convergence bins vary with cosmology.  In
the top panel of Figure~\ref{fig:Delta_Peak}, we show the difference
in peak numbers (averaged over the 200 realizations) in $3\times 3$
degree fields in convergence intervals $[\kappa,\kappa+\Delta\kappa]$,
with $\Delta\kappa=0.0161$ between the $w=-0.8$ (blue) and the
$\Lambda$CDM models, as well as $\Lambda$CDM and $w=-1.2$ (green).
The middle panel in the same figure shows the standard deviation in
each bin, and the bottom panel depicts the contribution to the
$\chi^2$ and thus to the distinction power, derived from each bin
(note that this bottom panel shows the unequal $\kappa$ bins that were
used in our $\chi^2$ calculation).

Based on the results shown in this figure, we identify three
categories of peaks.

\begin{itemize}

\item \textbf{High Peaks ($\kappa\gsim0.06$):} As is known from
previous work \citep{whiteprojection, Hamana:2003ts,Hennawi:2004ai}, a
significant fraction (and perhaps the majority) of these peaks are
caused by individual collapsed clusters, which can cause strong peaks
by themselves. These peaks are more numerous in cosmologies with more
negative $w$.  This is as expected: as mentioned above, the more
negative $w$ is, the later dark energy starts to dominate the energy
density, and the more time massive clusters had to form.

\item \textbf{Medium Peaks ($0.0\lsim \kappa\lsim0.06$):} The number
of these peaks significantly exceeds the expected number of clusters
that would cause, by themselves, peaks with a similar height.  We
examined the contributions of individual lens planes along the line of
sight (LOS) to the total convergence in these peaks. We found that for
the majority of these peaks, multiple planes contribute significantly,
and we therefore suspect these peaks are typically caused by
projection along the LOS.  In contrast with the high peaks, these
medium peaks are more numerous in cosmologies with less negative
$w$. This was unexpected.  Furthermore, we find that the difference in
the counts of these medium peaks dominates the cosmological
sensitivity.  Clarifying the physical nature of these peaks is
important, and will be presented in a follow-up paper
\cite{Yang:2009}.

\item \textbf{Low Peaks ($\kappa\lsim0.0$):} Similarly to the high
peaks, these peaks are more numerous in cosmologies with more negative
$w$. However, their number difference compared to their variance is
too low to make an appreciable contribution to the distinction between
the cosmologies (the number difference of the low peaks is only a
fifth of that of the medium peaks, yet their variance still about
half).  We speculate that these peaks typically reside in
``convergence voids'' (large areas of the map with very low values of
convergence).  We have indeed verified that the maps in cosmologies
with more negative $w$ show a larger fraction in voids, which gives
very low--amplitude peaks a larger area to reside in.

\end{itemize}

\subsection{Distinguishing $w\neq -1$ Models from $\Lambda$CDM}\label{Discerning}

We use the above difference in peak numbers in bins with different
values of $\kappa$ to quantify the statistical distinction between the
three models with $w=-1,-0.8$, and $-1.2$.  We find that the number of
peak height bins used has a minor effect, as long as their number is
kept between 4--9. Below this value, the binning is too coarse to
capture all the information of the rich structure in the top panel of
Figure~\ref{fig:Delta_Peak}, while higher numbers of bins show signs
of over-binning, especially when several source redshifts and
smoothing scales are combined (see Appendix \ref{Initial Conditions}).

As our fiducial set--up, we use a single source plane at $z_s=2$ and
$\theta_G=1$~arcmin smoothing, as in the previous section. Redshift
tomography and using different smoothing scales are addressed in
Sec.~\ref{Smoothing Scales} below. The $\chi^2$ for each of our
$3\times3$ degree convergence fields (with ellipticity noise and
smoothing), compared to the reference cosmology $\Lambda$CDM, is
calculated as described in \S~\ref{Comparison}. This leads to the
probability distribution of the $\chi^2$ shown in
Figure~\ref{fig:chi2}.

We can derive our statistical conclusions from these $\chi^2$
distributions.  For example, the mean of the $\chi^2$ distribution for
$w=-0.8$ lies at the 95\% tail of the $\chi^2$ distribution for
$w=-1$.  This suggests that the $2\sigma$ distinction from a single
$3\times3$ degree field corresponds to $\Delta w=0.2$.  The
corresponding result for the case of $w=-1.2$ is an exclusion at the
85\% confidence level.

\subsection{Extrapolation to an All--Sky Survey}

While we are limited by computational costs to studying single
$3\times3$ degree fields, the forthcoming surveys listed in
\S~\ref{Introduction} will cover much larger areas of the sky - up to
$\approx$ 20,000 degree$^2$.  Therefore, first we would like to
extrapolate our results to a larger field of view -- or equivalently,
translate it to the tighter $2\sigma$ sensitivity to $w$ expected from
larger fields.

We will use two methods of extrapolation, an optimistic and a
pessimistic one. In the optimistic case, we neglect large-scale
correlations and assume that, with increasing survey size, the
variance per area--squared in numbers of peaks just falls off as the
number of peaks, i.e.\ $\sigma^2_{N_{peak}}\propto N_{peak}$.  In the
pessimistic case, we extrapolate the scaling of the variance with
survey size from the scaling behavior in subfields of our $3\times3$
degree weak lensing maps.  We note that correlations on large scales
generally fall off faster than the Poisson--like extrapolation would
suggest \citep{Wang:2008hi}, and in reality, both cases may therefore
be conservative estimates.

\begin{figure}[htp]
\centering
\includegraphics[width=8 cm]{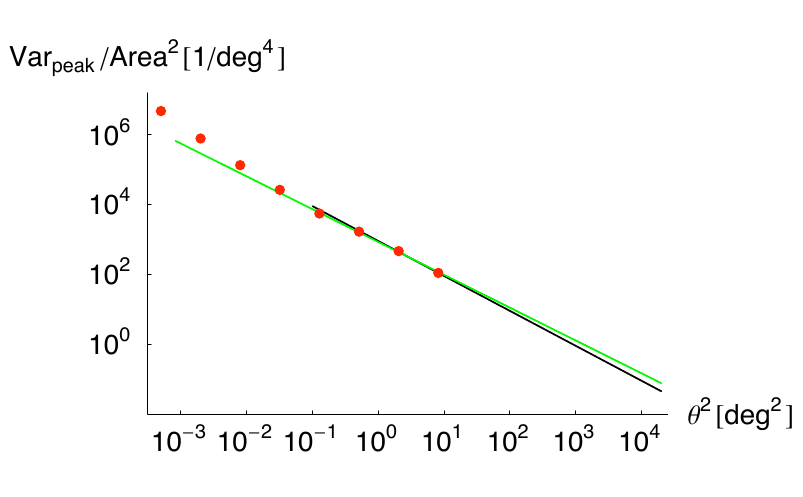}
\hfill
\caption[]{\textit{Variance (per area squared) in the total number of
convergence peaks, as function of survey area. The red dots denote the
variance as measured directly from our $3\times3$ degree convergence
maps and subfields therein, for the $w=-1$ cosmology, with sources at
redshift $z_s=2$ and 1 arcmin smoothing. The green line shows a
power--law fit to the four largest subfields and an extrapolation to a
20,000 degree$^2$ field. The black line depicts the slightly more
optimistic Poisson--like extrapolation
($\sigma^2_{N_{peak}}=N_{peak}$), which ignores long--range
correlations. The difference between these two cases causes a factor
of 1.3 difference in the extrapolated sensitivity to $w$ on large
scales.}}\label{fig:peak variance}

\end{figure}

We illustrate these two methods of extrapolations in
Figure~\ref{fig:peak variance}, where we plot the variance in peak
numbers for progressively larger fields, starting with a
$1.4\times1.4$ arcminute field and going all the way to $3\times3$
degrees (always obtained from our 200 pseudo-independent
realizations).

The optimistic extrapolation, shown by the black line in
Figure~\ref{fig:peak variance}, scales to a
$\sqrt{20,000/2.86^2}=50\times$ stronger sensitivity for a 20,000
degree$^2$ field of view.  The difference in the dark energy equation
of state parameters in our study is $\Delta w=0.2$, which we discern
at $2\sigma$. Therefore, roughly, a 20,000 degree$^2$ survey will be
able to discern a difference of $\Delta w=0.2/50=0.004$ at the same
$2\sigma$ level.  The somewhat more pessimistic case of extrapolating
from existing subfields, as depicted by the green line in
Figure~\ref{fig:peak variance}, yields a variance that is larger by a
factor of 1.3. Therefore, for this case, our estimated sensitivity $w$
from an LSST size survey is $\Delta w=0.0052$.

While this is an interesting level of sensitivity, one has to keep in
mind that this is a single--parameter constraint, and is not directly
comparable to marginalized constraints usually quoted in the
literature which fold in degeneracies with other parameters (see
\S~\ref{Other Constraints} below).

\subsection{Different Redshifts and Smoothing Scales}\label{Smoothing Scales}

In our fiducial set--up so far, we have used a single source plane
($z_s=2$) and smoothing scale ($\theta_G=1$~arcmin).  Here we examine
using other source planes and smoothing scales, including the
possibility of using multiple scales in combination.

\subsubsection{Single Redshifts and Smoothing Scale Combinations}

We have re--computed $\chi^2$ distributions, following the procedure
described above, for a variety of different smoothing scales
($\theta_G=1, 2, 3, 4, 5, 10$ arcmin) and source redshifts ($z_s=1,
1.5, 2$). We quote the results by indicating, in each case, the
fraction of the realizations of the $w=-0.8$ model which can be
excluded from the $w=-1$ cosmology at the 68\% (95\%) confidence
level.  These fractions are to be compared to the values of 32\% and
5\%, expected in the absence of any signal (i.e., for the same
cosmology, compared to itself).  These fractions are presented in
Table~1.  We list $\theta_G=0.5'$ in the table as well, but we caution
that, as discussed above, our simulations do not have sufficient
resolution to reproduce the convergence power spectrum on these
scales, and therefore these results may not be reliable.  Unless noted
otherwise, we exclude these values from the discussion in the rest of
this paper.

\begin{center}

\begin{tabular}{|c|c|c|c|} 
\hline
\% exclusion chance & & & \\
@ 68 (95)\% CL & $z_s=1$ & $z_s=1.5$ & $z_s=2$\\
\hline
$\theta_G=0.5'$ & 48 (13) \%  & 76 (32) \%  & 90 (46) \% \\
$\theta_G=1'$ & 56 (16) \%  & 68 (18) \%  & 84 (34) \%   \\
$\theta_G=2'$ & 53 (12) \%  & 66 (13) \%  & 50 (18) \%   \\
$\theta_G=3'$ & 41 (10) \%  & 48 (14) \%  & 40 (6) \%   \\
$\theta_G=4'$ & 39 (6) \%  & 32 (7) \%  & 37 (12) \%   \\
$\theta_G=5'$ & 38 (6) \%  & 28 (8) \%  & 39 (10) \%   \\
$\theta_G=10'$ & 26 (2) \%  & 28 (5) \%  & 25 (4) \%   \\
\hline
\end{tabular}

\end{center}

Table 1: \textit{The probability of distinguishing the $w=-0.8$ model
from $\Lambda$CDM, based on peak counts in a single $3\times3$ degree
field, at the 68\% (95\%) confidence level for various source
redshifts $z_s$ and smoothing scales $\theta_G$, using 8 peak height
bins.}

\vspace{0.4cm}

There are several trends worth noting in Table~1. First, using source
galaxies at high redshift ($z_s=2$) provides substantial advantage
over using sources at redshift $z_s=1.5$ or even $z_s=1$ (recall that
the surface density is fixed in each case to be 15 arcmin$^{-2}$).
Dark energy constraints from Type Ia supernovae, and from most other
techniques also benefit from deep surveys and high redshifts. For weak
lensing, going to higher redshifts is particularly important, as the
best lensing efficiency comes from approximately halfway in distance
between sources and observer.

Second, as is well known, weak lensing derives most of its power from
small scales, and indeed, the smaller the smoothing scale in our maps,
the stronger the distinction. This is true down to the smallest
angular scales that we can reliable simulate, $\theta_G\approx 1$.
Smoothing scales larger than $\theta_G=2'$ do not provide very strong
constraints by themselves, but may still be useful when combined with
smaller ones (especially for real surveys with much larger fields of
view -- the number of peaks at these large scales becomes very small
for our small $3\times3$ degree fields).

\subsubsection{Correlations}

There are several types of correlations that would be useful to
investigate: i.e., correlations in the number of peaks in different
height bins, identified for different source redshifts, and with
different smoothing scales, as well as the cross-correlations between
these quantities.  In order to study the effect of these correlations,
we computed the $\chi^2$ distributions from covariance matrices in
which only diagonal elements, or diagonal blocks were retained, and
the cross-terms outside these were set to zero.

We find that the peak height bins are correlated. Neglecting all
correlations, by retaining only the diagonal elements of the
covariance matrix, lead to $\chi^2$ histograms that resemble $\chi^2$
distributions with fewer degrees of freedom, and an excessive amount
of outliers, both typical signs that the dataset contains more
correlations than is assumed in the analysis. In particular, as
expected, the $\chi^2$ distributions (analogous to those shown in
Figure~\ref{fig:chi2}) develop wider tails toward high $\chi^2$
values, with their maximum moving to somewhat smaller values. The net
result is that the overlap of the $\chi^2$ distributions between a
pair of cosmologies is enhanced, and the distinction between the
cosmologies is degraded.  Table~2 illustrates this in numbers.

\begin{center}

\begin{tabular}{|c|c|c|c|} 
\hline
\% exclusion chance & & & \\
@ 68 (95)\% CL & $z_s=1$ & $z_s=1.5$ & $z_s=2$\\
\hline
$\theta_G=0.5'$ & 32 (1) \%  & 46 (2) \%  & 52 (2) \% \\
$\theta_G=1'$ & 34 (2) \%  & 54 (2) \%  & 65 (8) \%  \\
$\theta_G=2'$ & 40 (3) \%  & 51 (5) \%  & 44 (8) \%   \\
$\theta_G=3'$ & 39 (2) \%  & 44 (6) \%  & 38 (4) \%   \\
$\theta_G=4'$ & 36 (2) \%  & 30 (6) \%  & 32 (5) \%   \\
$\theta_G=5'$ & 32 (4) \%  & 32 (3) \%  & 36 (6) \%   \\
$\theta_G=10'$ & 24 (2) \%  & 32 (4) \%  & 26 (3) \%   \\
\hline
\end{tabular}

\end{center}

Table 2: \textit{Same as Table 1, but with correlations between peak
height bins neglected. The likelihood of distinguishing $w=-0.8$ from
$\Lambda$CDM are much lower, and approach the no-significance values
of 32 (5) \%, in particular for the 95\% confidence level numbers,
indicating a large number of outliers.}

\vspace{0.4cm}

We also attempted to study cross-correlations between different
redshifts and smoothing scales. When studying several redshifts and
smoothing scales simultaneously, the number of elements in the
covariance matrix grows quickly.  Because of the low number of
realizations and the corresponding underestimation of the variance,
results obtained by using more than $\approx 9$ bins become unreliable
(Appendix \ref{Initial Conditions}).  Unfortunately, this prohibits us
from studying the effect of correlations between redshifts and
smoothing scales reliably: the number of different peak height bins
still allowed becomes too small.  In particular, when multiple source
redshifts or smoothing scales are simultaneously included, only 2--3
peak heights are allowed by the low number of realizations. We found
that such a low number of convergence bins does not sufficiently
capture the peak number difference distributions in
Figure~\ref{fig:Delta_Peak}, and no meaningful constraints are
possible.  We defer a study of these cross--correlations to a
follow--up paper, with a larger number of independent realizations of
each cosmology.

\subsection{Comparison to Other Forecasts for LSST}\label{Other Constraints}

It is useful to place our results in the context of constraints
expected to be available from other uses of large WL maps.  Three
methods we will briefly mention here are (i) the shear power spectrum,
(ii) cluster number counts, and (iii) the one--point function of the
convergence PDF.

(i) From an 11-parameter fit to the tomographic shear power spectrum
from LSST and constraints from Planck, \cite{S&K04} obtain
$\sigma(w_0)=0.06, \sigma(w_a)=0.09$.

(ii) From a 7-parameter fit to the number counts of $\approx$ 200,000
shear-selected clusters, combined with Planck constraints, \cite{SW04}
obtain $\sigma(w_0)=0.04, \sigma(w_a)=0.09$.

(iii) Using the fractional area statistic of areas of high
convergence, \cite{Wang:2008hi} found $\sigma(w_0)=0.043,
\sigma(w_a)=0.11$ for LSST and priors from Planck.

The above are all $1\sigma$ constraints, marginalized over a large
number of parameters. We do not attempt a fair comparison of the three
methods, which would require that identical assumptions are made about
the survey specifications and about the set of free parameters (see
\cite{DETF} for a similar exercise covering i. and ii.).
Nevertheless, these numbers demonstrate, first, that the statistical
information contained in the non--linear features in WL maps --
captured by (ii) and (iii) -- is at least roughly comparable to the
information encoded in the power spectrum in (i).  Second,
our single--parameter sensitivity to $w$ can be compared meaningfully
to those from clusters counts.  In particular, as mentioned in
\S~\ref{Peak Numbers}, the counts of clusters above the $4.5\kappa_G$
convergence threshold differs in the $w=-0.8$ and $w=-1$ cosmologies
by $\approx 1.5\sigma$.  We find that the number of convergence peaks
above the same threshold differs in these two cosmologies at a
similar, $\approx 1\sigma$ significance level. We find further that
considering lower height peaks, with convergence well below this
threshold, significantly improves the distinction.  While follow-up
work, with simulations that address degeneracies between $w$,
$\sigma_8$, and other parameters is needed, these results suggest that
the parameter sensitivity from WL peak counts will be competitive with
those from the above three methods.

As mentioned in \S~\ref{Introduction}, in a recent preprint,
\cite{Dietrich:2009jq} independently addressed the dependence of WL
peaks on the background cosmology. \cite{Dietrich:2009jq} used the
same simulation box size as in our work, but a smaller number of
particles ($256^3$, rather than $512^3$), which allowed them to sample
a 2D parameter space and initial conditions, at the cost of mass and
angular resolution. They compute the aperture mass for the peaks as a
weighted integral of the tangential shear, rather than the
convergence. Despite these differences, the conclusions of the two
papers appear to be remarkably similar. In particular, if we attribute
the distinction we find between our cosmologies with different $w$
values entirely due to the differences in $\sigma_8$ in these
cosmologies, we find a very similar sensitivity to the latter
parameter as their Figure 3: $0.73\lsim\sigma_8\lsim 0.83$ (for a
fixed $\Omega_m=0.26$). We find, additionally, a subdominant type of
peaks -- low--amplitude peaks, residing in the voids -- the
distinction between the cosmologies is however dominated by medium
peaks, with a strong contribution from high peaks.

\section{Summary}
\label{Summary}

We propose the new method of using the counts of peaks, identified in
weak gravitational lensing maps, to constrain dark energy and other
cosmological parameters. The method makes no reference to cluster
counts or their mass function, and so, by definition, is free from
selection effects inherent in methods that target clusters.

We used N--body simulations to create theoretical convergence maps in
three cosmologies with $w=-0.8$, $w=-1$, and $w=-1.2$. We identify
three different peak types: high, medium, and low peaks, whose numbers
depend differently on cosmology. The high ($\kappa\gsim 0.6$) and low
($\kappa\lsim 0$) peaks both become more numerous with increasingly
negative $w$, while the opposite trend is true for the medium
peaks. For the high peaks, this dependence confirms the expectation
that when $w$ is more negative, dark energy begins to dominate later,
and more time is allowed for the growth of non--linear structures.
The opposite behavior of the medium peaks is a surprise - which is all
the more important since we find that these peaks dominate the
distinction between the cosmologies. The majority of these peaks are
not typically caused by individual collapsed clusters, and are more
likely due to projection of large scale structure along the line of
sight. The low peaks form a subdominant contribution. These peaks are
typically due to random ellipticity noise, and we speculate that the
larger surface area of voids in cosmologies with more negative $w$,
which we have detected in the maps, offers more opportunities for
these small fluctuations to arise.

We have shown that based on these peak counts, cosmologies with
$w=-0.8$ and $w=-1.2$ can be distinguished from $\Lambda$CDM with a
single $3\times3$ degree convergence field at a 95\% and 85\%
confidence level, respectively. Extrapolating these results to the
20,000 degree$^2$ field of view of a large survey, such as LSST, we
obtain a projected sensitivity to $w$ of $\Delta w=0.004$--$0.0052$,
depending whether the extrapolation is based on the variance measured
in subfields of our maps, or whether a simple Poisson--like scaling is
used.

In this work, we have explored three constant values of the dark
energy equation of state, $w=-0.8$, $w=-1$, and $w=-1.2$, representing
variations around the best fit $\Lambda$CDM model to the 5-year WMAP
data.  The most significant concern is that there will be strong
degeneracies between $w$ and other parameters; our results may already
be driven primarily by the $\sigma_8$ sensitivity in the number of
peaks.  In a followup paper, we will study such degeneracies; it will
also be necessary to perform a larger number of realizations of each
cosmology, in order to study the utility of combining several source
galaxy redshifts and to use information possibly contained in the
angular size distribution of the peaks. In a separate publication
\cite{Yang:2009}, we also plan to investigate in more detail the
physical origin of the three kinds of peaks detected in this work.

\begin{acknowledgments}

  We would like express our deep thanks to Lam Hui for numerous
  helpful discussions, and to Greg Bryan, Francesco Pace, and Matthias
  Bartelmann and his group for invaluable help during code
  development. We are also grateful to Christof Wetterich for kindly
  supporting an extended visit by JK at the University of Heidelberg.
  We also thank Puneet Batra, Wenjuan Fang, Eugene Lim and Sarah
  Shandera for useful discussions about statistics, CAMB, and lensing,
  and Volker Springel for his help with GADGET-2 and for providing us
  with his parallelized initial conditions generator.  JK is supported
  by ISCAP and the Columbia Academic Quality Fund.  This work was
  supported in part by the Pol\'anyi Program of the Hungarian National
  Office for Research and Technology (NKTH), by NSF grant
  AST-05-07161, by the U.S. Department of Energy under contract
  No. DE-AC02-98CH10886, and by the Initiatives in Science and
  Engineering (ISE) program at Columbia University.  The computational
  work for this paper was performed at the LSST Cluster at Brookhaven
  National Laboratory and with the NSF TeraGrid advanced computing
  resources provided by NCSA.

\end{acknowledgments}

\section{Appendix: Initial Conditions and the Number of Realizations}\label{Initial Conditions}

In this Appendix, we show our results as a function of the number of
peak height bins used, and justify the choice for the number of bins
we made in the body of the paper.

Ideally -- in the limit that the covariance matrix was perfectly
computed -- the distinguishing power would asymptotically approach a
constant value as one increases the number of bins, at which point
using more bins will not provide any more benefit.  Unfortunately, it
is computationally expensive to generate a large ensemble of initial
conditions from which to construct several strictly independent
realizations of a given cosmology (a single realization takes
approximately 4800 CPU hours, i.e. ca. 2.5 days on 80 CPUs). This,
together with the storage required for the large amount of data
produced in each run, ultimately limits the number of bins we can
utilize.

In particular, using too few realizations, together with too many
bins, will generally result in an underestimate of the variance, and
an artificial boost in the $\chi^2$ values.  The possible concern,
then, is that the differences we find in the convergence maps between
two cosmologies may be caused by such an underestimation of the
variance.

In order to guard against this possibility, and to verify that the
differences in the convergence maps are caused by genuine differences
in cosmology, we generated a second set of independent initial
conditions for the cosmologies with $w=-1$ and with $w=-0.8$.  The
simple Ansatz we adopt for the maximum number of bins that can be
safely used, without underestimating variances, is the following: when
two different realizations of the same $\Lambda$CDM cosmology are
compared, we should recover the $\chi^2$ distribution, i.e., we must
not be able to rule out the fiducial model itself.

\begin{widetext}

\begin{figure}[htp]
\centering
\includegraphics[width=8 cm]{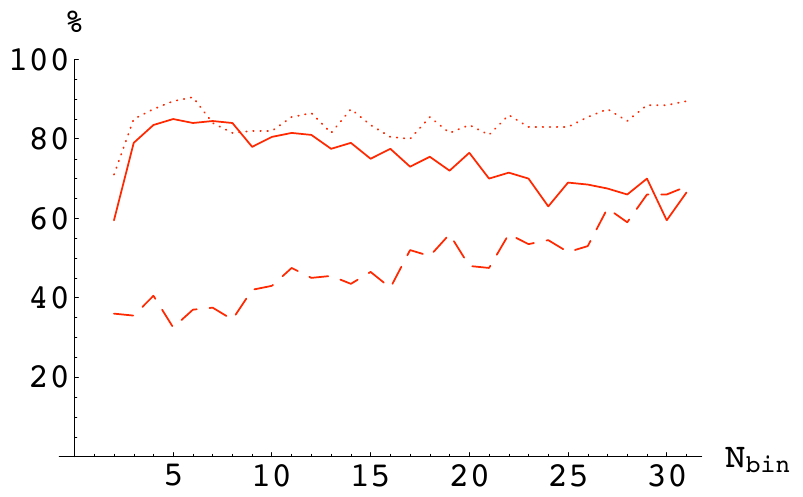} \includegraphics[width=8 cm]{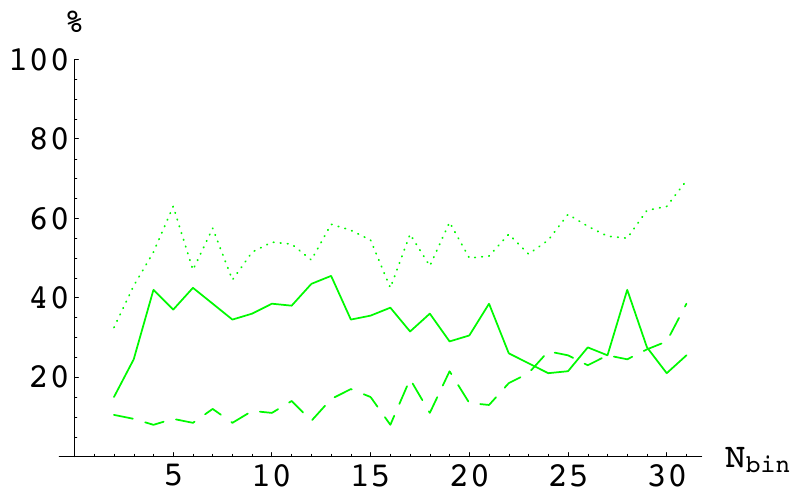}
\hfill
\caption[]{\textit{The solid curves show the percentage of maps in the
    $w=-0.8$ cosmology which can be excluded at the 68\% (left panel)
    and 95\% (right panel) confidence level from $\Lambda$CDM.  In
    these comparisons, the maps in the $w=-0.8$ cosmology were
    generated from initial condition A, which were compared to the
    expected average in the $\Lambda$CDM cosmology with initial
    condition B; the covariance matrix was calculated from
    $\Lambda$CDM with initial condition A. The dotted curves show the
    same quantities, except that the maps in the $w=-0.8$ model were
    taken from initial condition B, and those from $\Lambda$CDM with
    initial conditions A.  The dashed curves again show the same
    quantities, except maps from the $\Lambda$CDM model (with initial
    condition B) were compared against $\Lambda$CDM with initial
    condition A and covariance matrix calculated from initial
    condition A. If the covariance matrix was correctly estimated, the
    latter curve should be flat, whereas it is begin to rise
    systematically for 10 or more bins. Based on this result, we
    identify 9 as the maximum number of bins that can be safely used
    without underestimating variances. }}\label{fig:Bindependence}

\end{figure}

\end{widetext}

In Figure~\ref{fig:Bindependence}, we show the percentage of the
realizations of the $w=-0.8$ test cosmology that lies beyond the 68\%
(left panel) and 95\% (right panel) tail of the $\chi^2$ distribution
(solid curves), together with the same quantities for the fiducial
$\Lambda$CDM cosmology itself (dashed curves). Our Ansatz requires
that the dashed curves correspond to the confidence level at which the
exclusion is to be achieved (e.g. no more than 32 / 5\% of maps
excluded at the 68 / 95\% confidence level). We see that this
condition is satisfied when 4-9 bins are used, but for 10 or more
bins, there is a steadily rising difference between the two
realizations of the $\Lambda$CDM model. We therefore take 9 bins as
the maximum acceptable total number of bins.

Figure~\ref{fig:Bindependence} also shows that the distinction between
$w=-0.8$ and the $\Lambda$CDM model is relatively steady for 4-9 bins,
and is insensitive to the choice of the realization of the initial
condition.  However, for a larger number of bins, the distinction
starts to be either less significant (when different realizations of
the initial conditions are used in the two cosmologies; solid curves),
or begin a modest rise in significance (when the same realizations are
used; dotted curves).

While the above analysis is based only on the number of convergence
bins, we find similar conclusions when the bins include multiple
source galaxy redshifts and/or multiple smoothing scales.  We conclude
that the accuracy of our covariance matrix limits us to a total of 9
bins.

\vfill

\begin{thebibliography}{4}

\bibitem[{Albrecht} et~al.(2006){Albrecht}, {Bernstein}, {Cahn}, {Freedman},
  {Hewitt}, {Hu}, {Huth}, {Kamionkowski}, {Kolb}, {Knox}, {Mather}, {Staggs},
  and {Suntzeff}]{DETF}
{Albrecht}, A., G.~{Bernstein}, R.~{Cahn}, W.~L. {Freedman}, J.~{Hewitt},
  W.~{Hu}, J.~{Huth}, M.~{Kamionkowski}, E.~W. {Kolb}, L.~{Knox}, J.~C.
  {Mather}, S.~{Staggs}, and N.~B. {Suntzeff}, 2006: {Report of the Dark Energy
  Task Force}. {\em e-prints astro-ph/0609591} (2006).

\bibitem{RefReview} A.~{Refregier}, 
    Ann.\ Rev.\ Astron.\ Astrophys.\  {\bf 41}, 645 (2003).
  [arXiv:astro-ph/0307212].
  
\bibitem{Kaiser:1999kka}
  N.~Kaiser, J.~L.~Tonry and G.~A.~Luppino,
  arXiv:astro-ph/9912181 (1999).
  
  \bibitem{Tyson:2002nh}
  J.~A.~Tyson, D.~M.~Wittman, J.~F.~Hennawi and D.~N.~Spergel,
  Nucl.\ Phys.\ Proc.\ Suppl.\  {\bf 124}, 21 (2003).
  [arXiv:astro-ph/0209632].
  
 \bibitem{Tyson2002b}
 J.~A.~Tyson and the LSST Collaboration,
  Proc.\ SPIE Int.\ Soc.\ Opt.\ Eng.\ {\bf 4836}, 10--20 (2002).
  See also http://www.lsst.org
  

\bibitem[{Kaiser}(1992)]{Kaiser92}
N.~Kaiser,
  ``Weak Gravitational Lensing Of Distant Galaxies,''

\bibitem[{Jain} and {Seljak}(1997)]{JS97}
B.~Jain and U.~Seljak,
  Astrophys.\ J.\  {\bf 484}, 560 (1997).

\bibitem[{Hu}(1999)]{WH99}
W.~Hu,
  Astrophys.\ J.\  {\bf 522}, L21 (1999).

  
  \bibitem[{Hu}(2002)]{WH02}
  W.~Hu,
  Phys.\ Rev.\  D {\bf 66}, 083515 (2002).
  
\bibitem[{Huterer}(2002)]{DH02}
D.~Huterer,
  Phys.\ Rev.\  D {\bf 65}, 063001 (2002).
  
  \bibitem[{Refregier} et~al.(2004){Refregier}, {Massey}, {Rhodes}, {Ellis},
  {Albert}, {Bacon}, {Bernstein}, {McKay}, and {Perlmutter}]{AR04}
A.~Refregier {\it et al.},
  Astron.\ J.\  {\bf 127}, 3102 (2004).
  
  
  \bibitem[{Abazajian} and {Dodelson}(2003)]{A&D03}
  K.~N.~Abazajian and S.~Dodelson,
  Phys.\ Rev.\ Lett.\  {\bf 91}, 041301 (2003).
  
\bibitem[{Takada} and {Jain}(2004)]{T&J04}
M.~Takada and B.~Jain,
  Mon.\ Not.\ Roy.\ Astron.\ Soc.\  {\bf 348}, 897 (2004).


\bibitem[{Song} and {Knox}(2004)]{S&K04}
  Y.~S.~Song and L.~Knox,
  Phys.\ Rev.\ D {\bf 70}, 063510 (2004).
  
  
\bibitem[{White} et~al.(2002){White}, {van Waerbeke}, and
  {Mackey}]{whiteprojection}
  M.~J.~.~White, L.~van Waerbeke and J.~Mackey,
  Astrophys.\ J.\  {\bf 575}, 640 (2002).
  
  
\bibitem{Hamana:2003ts}
  T.~Hamana, M.~Takada and N.~Yoshida,
  Mon.\ Not.\ Roy.\ Astron.\ Soc.\  {\bf 350}, 893 (2004).

\bibitem{Hennawi:2004ai}
  J.~F.~Hennawi and D.~N.~Spergel,
  Astrophys.\ J.\ {\bf 624}, 59--79 (2005).


\bibitem[{Wang} and {Steinhardt}(1998)]{W&S98}
L.~M.~Wang and P.~J.~Steinhardt,
  Astrophys.\ J.\  {\bf 508}, 483 (1998).

\bibitem[{Haiman} et~al.(2001){Haiman}, {Mohr}, and {Holder}]{H&M&H01}
Z.~Haiman, J.~J.~Mohr and G.~P.~Holder,
  Astrophys.\ J.\  {\bf 553}, 545 (2001).

  
  \bibitem[{Weller} et~al.(2002){Weller}, {Battye}, and {Kneissl}]{W&B&K02}
  J.~Weller, R.~Battye and R.~Kneissl,
  Phys.\ Rev.\ Lett.\  {\bf 88}, 231301 (2002).
  
 
\bibitem{SW04}
 S.~Wang, J.~Khoury, Z.~Haiman and M.~May,
  Phys.\ Rev.\  D {\bf 70}, 123008 (2004).
   
\bibitem{Marian:2008fd}
L.~Marian, R.~E.~Smith and G.~M.~Bernstein,
  Astrophys.\ J.\  {\bf 698}, L33 (2009).

\bibitem{Chanta}
T. Chantavat, C. Gordon, and J. Silk,
  Phys. Rev. D, {\bf 79}, 083508, (2009).

\bibitem{Fang:2006dt}
  W.~J.~Fang and Z.~Haiman,
  Phys.\ Rev.\  D {\bf 75}, 043010 (2007).

 \bibitem{Takada:2007}
  M.~Takada, and S.~Bridle,
  New Journal of Physics, {\bf 9}, 446 (2007).


\bibitem[{Marian} and {Bernstein}(2006)]{M&B06}
L.~Marian and G.~M.~Bernstein,
  Phys.\ Rev.\  D {\bf 73}, 123525 (2006).


\bibitem{Schirmer:2006ud}
  M.~Schirmer, T.~Erben, M.~Hetterscheidt and P.~Schneider,
  Astron.\ Astrophys.\  {\bf 462}, 875 (2007).

\bibitem{Dietrich:2007gi}
  J.~P.~Dietrich, T.~Erben, G.~Lamer, P.~Schneider, A.~Schwope, J.~Hartlap and M.~Maturi,
  Astron.\ Astrophys.\ {\bf 470}, 821(2007).


\bibitem[{Jain} and {Van Waerbeke}(2000)]{BJ&VW00}
B.~Jain and L.~V.~Van Waerbeke,
  Astrophys.\ J.\  {\bf 530}, L1 (2000).
  
\bibitem{Wang:2008hi}
  S.~Wang, Z.~Haiman and M.~May,
  Astrophys.\ J.\  {\bf 691}, 547 (2009).

\bibitem{Dietrich:2009jq}
  J.~P.~Dietrich and J.~Hartlap,
  e-print arXiv:0906.3512.

\bibitem{Springel:2005mi}
  V.~Springel,
  Mon.\ Not.\ Roy.\ Astron.\ Soc.\  {\bf 364}, 1105 (2005).
  
  
\bibitem{Lewis:1999bs}
  A.~Lewis, A.~Challinor and A.~Lasenby,
  Astrophys.\ J.\  {\bf 538}, 473 (2000).
  
  
\bibitem{Seljak:1996is}
  U.~Seljak and M.~Zaldarriaga,
  Astrophys.\ J.\  {\bf 469}, 437 (1996).
  
\bibitem{Hockney-Eastwood}
R.~W.~Hockney, J.~W.~Eastwood,
Adam Hilger, Bristol  (1988).

\bibitem{FITS}
D.~C.~Wells, E.~W.~Greisen and R.~H.~Harten,
Astronomy and Astrophysics Supplement Series {\bf 44}, 363-370 (1981).

\bibitem{Hamana:2001vz}
  T.~Hamana and Y.~Mellier,
  Mon.\ Not.\ Roy.\ Astron.\ Soc.\  {\bf 327}, 169 (2001).


\bibitem{Dodelson:2005ir}
  S.~Dodelson, C.~Shapiro and M.~J.~.~White,
  Phys.\ Rev.\  D {\bf 73}, 023009 (2006).
  
  
\bibitem{VanWaerbeke:1999wv}
  L.~Van Waerbeke,
  Mon.\ Not.\ Roy.\ Astron.\ Soc.\  {\bf 313}, 524 (2000).
    

\bibitem{Limber}
  D.~N.~Limber,
  Astrophys.\ J.\ {\bf 117}, 134 (1953).


\bibitem{Smith} {Smith}, R.~E., J.~A. {Peacock}, A.~{Jenkins},
S.~D.~M. {White}, C.~S. {Frenk}, F.~R. {Pearce}, P.~A. {Thomas},
G.~{Efstathiou}, and H.~M.~P. {Couchman},
  MNRAS, {\bf 341}, 1311 (2003).

\bibitem{Jenkins} A. Jenkins {\it et al.}, Mon. Not. R. Astron. Soc. {\bf 321}, 372 (2001).
  
\bibitem{NFW}
  J.~F.~Navarro, C.~S.~Frenk and S.~D.~M.~White,
  Astrophys.\ J.\  {\bf 490}, 493 (1997).

\bibitem{HuKravtsov} W. Hu, \& A. V. Kravtsov, Astrophys. J. {\bf 584}, 702 (2003).

\bibitem{Yang:2009}
 X.~Yang, J.~.M.~Kratochvil, Z.~Haiman, M.~May, 
 in preparation.

\bibitem{Pires:2009ar}
  S.~Pires, J.~L.~Starck, A.~Amara, A.~Refregier and R.~Teyssier,
  e-print arXiv:0904.2995



\end{thebibliography}
\end{document}